\documentstyle[aps]{revtex}
\begin{document}
\draft
\title{Nonequilibrium total dielectric function approach to the electron
Boltzmann equation
for inelastic scattering in doped polar
semiconductors}
\author{B.A. Sanborn}
\address{Semiconductor Electronics Division \\
National Institute of Standards and Technology, Gaithersburg, MD\,\, 20899}
%\date{\today}
\maketitle
\begin{abstract}
This paper describes a simple and general method for deriving the
inelastic collision term in the electron Boltzmann
equation for scattering from a coupled electron-phonon system, and
applies the method to the case of doped polar semiconductors.
In the Born approximation, the inelastic differential scattering rate
$W^{inel}$
can be expressed in terms of the nonequilibrium total dynamic dielectric
function, which includes both electronic and lattice contributions.
Within the random-phase approximation
$W^{inel}$ separates into two components: an electron-electron
interaction containing the nonequilibrium distribution function for
excitations of the electron gas,
and a Fr\"{o}hlich interaction including the phonon
distribution function and
self-energy due to polarization of the electrons. Each
of these two interactions is screened by only the electronic part of the total
dielectric function which contains the high frequency dielectric constant,
unlike commonly used expressions which contain the static dielectric
constant. Detailed balance between plasmons and electron-hole pairs
in steady state is used to eliminate the nonequilibrium plasmon
distribution from the Boltzmann equation, resulting in a
dynamically screened
electron-electron collision term.
The phonon self-energy
modifies the longitudinal optical phonon dispersion
so that two hybrid normal modes contribute to the
electron-phonon collision term.
\end{abstract}
%\pacs{72.10.Bg, 72.10.-d, 72.20.Dp, 72.20.Fr, 72.80.Ey}
\section{Introduction}
The electron Boltzmann equation for inelastic scattering in solids is usually
established either by making approximations to more general equations derived
with
powerful nonequilibrium Green's function techniques\cite{KAD,HOL,DUB,SPI},
or simply
by ansatz with the aid of analogy to simpler physical systems. An alternative
method described here proceeds by using time-dependent perturbation theory
to determine the
inelastic collision term from the total dynamic dielectric
function for the nonequilibrium
coupled electron-phonon system.
This paper treats the case of a system of polar optical
phonons in the presence of conduction electrons
and Coulomb interactions, though the method is more
generally applicable.
Except for the nonequilibrium character of the Boltzmann problem,
the approach is similar
to Van Hove's
correlation function
method\cite{VAN} applied to doped polar semiconductors\cite{KIM}
to determine the inelastic lifetime
of a quasiparticle excitation of an equilibrium system.
In addition to
providing physical insight, the total dielectric function approach is a
systematic way to include scattering against coupled electron-phonon modes
in the Boltzmann equation, so that the influence of mode coupling on
mobility can be studied. Numerical results for mobility in n-type GaAs are
presented in reference \cite{GAAS}.

In the Born approximation, the interaction between a
conduction electron and a coupled electron-phonon system can be represented
as an effective electron-electron interaction screened by the total
dynamic longitudinal dielectric function,
$\epsilon_{T}({\bf q},\omega)$, which includes contributions from
both electrons and phonons\cite{PENN,ALLEN,MAHAN}.
In this context, $\epsilon _{T}$ describes the linear response of the
electron-phonon gas not to an externally applied potential, but to
the {\em internal} potential due to the probe electron plus the
induced density
fluctuations of the gas itself.
The effective interaction has the form of a total density autocorrelation
function.
Unlike the lifetime problem\cite{KIM},
the entire electron-phonon gas is away from
equilibrium in the dc transport case, and the
correlation function must be evaluated for the {\em nonequilibrium} ensemble.
It follows that the effective electron-electron
interaction is screened by the nonequilibrium total dielectric function
in this case.
In fact, taking account of the nonequilibrium
character of the interaction is essential to obtaining the correct form of the
inelastic collision term, which contains nonequilibrium distribution
functions for the system's excitations.

The effective electron-electron interaction can be separated into
a purely electron-electron term and a screened electron-phonon
interaction.
In the random-phase approximation (RPA),
the electron-phonon part
contains a phonon self-energy which arises from the polarization
of the electron gas\cite{MAHAN}.
The self-energy correction modifies the longitudinal optical (LO) phonon
dispersion
in doped polar semiconductors,
producing two hybrid normal modes with
phonon strength in both.
This paper
determines the inelastic differential
scattering rate $W^{inel}$ for the nonequilibrium case
in the RPA Born approximation.
The result consists of an electron-electron component
containing the nonequilibrium distribution function for
excitations of the electron gas, and an electron-phonon component
containing the phonon self-energy and distribution function.
Each of these two interactions is screened by {\em only} the electronic part of
the total dielectric function, which contains the high-frequency dielectric
constant $\epsilon_{\infty}$.
Thus, the RPA Born result differs from commonly used expressions for
screened interactions in doped polar semiconductors
which contain the static dielectric constant $\epsilon_{0}$. Examples using
$\epsilon_{0}$ rather than $\epsilon_{\infty}$ for n-type GaAs include a
calculation of dynamically screened
electron-electron scattering\cite{LUG},
as well as static screening approximations for
electron-LO phonon scattering\cite{LOW}
and electron-electron scattering\cite{CHATT}.
For more strongly polar doped semiconductors (where $\epsilon_{0} \gg
\epsilon_{
\infty}$), it is essential to treat this aspect of the screening problem
correctly.

The collective modes have a finite lifetime for decay
into electron-hole
pairs. This Landau damping process provides an indirect mechanism
for single-particle electron-electron scattering.
For the electronic system
in steady state, Landau damping creates a
detailed balance relation between plasmons and electron-hole pairs
which can be used to eliminate
the nonequilibrium plasmon distribution from the
Boltzmann equation
in favor of the single-particle electron distribution.
In the approximation of no collision damping,
the result is an
electron-electron collision term screened in the dynamic RPA,
containing
the effects of electron-plasmon scattering.
For direct gap doped semiconductors where Umklapp processes are negligible,
this collision term conserves total electron momentum and cannot by itself
degrade an electrical current if energy surfaces are spherical. However, by
rearranging the electron momentum distribution, electron-plasmon and
electron-electron scattering can have a significant indirect effect on
the outcomes of other scattering processes.

The low-energy hybrid mode arising from the plasmon-phonon coupling in doped
polar semiconductors gives increased low-energy electron-phonon scattering
strength compared to the uncoupled case. Numerical calculations including the
plasmon-phonon coupling have given shorter inelastic lifetimes\cite{KIM,JAL}
and enhanced hot electron energy relaxation\cite{DASSAR,LEI}, especially
at low temperatures and low doping. A simple way
to treat the coupled modes is to use the plasmon-pole
approximation\cite{PP,PPSE} on the phonon self-energy.
This procedure is used here to include scattering against the hybrid
modes in the electron-phonon collision term so that mode coupling can be
treated
with the Boltzmann equation in a computationally efficient
manner.

This paper is organized as follows.
In section II, $W^{inel}$ is derived
within the Born approximation
in terms of the total nonequilibrium dielectric function
for the coupled electron-phonon
system. The RPA is invoked to separate $W^{inel}$ into electron-electron
 and electron-phonon components.
Section III
deals with the electron Boltzmann equation by eliminating the nonequilibrium
plasmon distribution from the electron-electron collision term.
The electron-phonon collision term is simplified by applying the plasmon-pole
approximation to the phonon self-energy.

\section{Inelastic differential scattering rate}
\subsection{Electron-phonon gas model}
Consider a probe electron
scattering from an interacting
gas of electrons and polar optical phonons.
Interactions with impurities and acoustic phonons can be added by hand later.
The probe electron is treated as a
quasiparticle with states $|{\bf k}\rangle$, effective
mass $m^{*}$, and dispersion relation $E_{k}=\hbar^{2}k^{2}/2m^{*}$.
The Boltzmann equation relies on the validity of
the quasiparticle description by assuming that the energy-momentum relation
remains well defined when collisions are included.
Interactions with the electron-phonon gas produce transitions between
the quasiparticle states, but do not broaden them significantly.

For direct-gap polar semiconductors, local field effects are unimportant
so that electrons couple only to the macroscopic electric field set up by
the LO phonons, and do not interact with the transverse optical (TO) phonons.
The  Hamiltonian for the electron-LO phonon gas is\cite{MAHAN}
\begin{equation}
\tilde{H}_{g}=\sum _{{\bf k},\sigma}{\cal E}_{k}c^{\dag}_{{\bf k},\sigma}
c_{{\bf k},\sigma}
+\sum_{{\bf q}}\hbar \omega_{TO}
(a^{\dag}_{{\bf q}}
a_{{\bf q}}+\frac{1}{2})
+\frac{1}{2}\sum _{{\bf q}\neq 0}v_{q}^{\infty}\left[\rho _{T}^{\dag}
({\bf q})\rho_{T}({\bf q})-N\right]\,, \label{Hg}
\end{equation}
where
\begin{equation}
v_{q}^{\infty} = \frac{v_{q}}{\epsilon_{\infty}}
= \frac{4 \pi e^{2}}{q^{2}\epsilon_{\infty}} \nonumber \,,
\end{equation}
and where the electron and LO phonon annihilation operators are denoted by
$c_{{\bf k}}$ and $a_{{\bf q}}$, respectively, $\sigma$ labels the electron
spin,
and $\omega _{TO}$
is the TO phonon frequency.
The relevant phonon frequency in $\tilde{H}_{g}$ is the bare one,
$\omega _{TO}$.
The ion-ion Coulomb interactions
explicitly included in $\tilde{H}_{g}$ shift $\omega _{TO}$ to
the longitudinal optical frequency $\omega _{LO}$, so
that only $\omega _{LO}$ appears in expressions for observables\cite{MAHAN}.
The first and second sums in $\tilde{H}_{g}$ describe the noninteracting
systems
of electrons and phonons, respectively. The last sum includes
electron-electron,
electron-phonon, and ion-ion Coulomb interactions.
The total density operator $\rho _{T}$ is the summation of the density
operators of the electrons and LO phonons,
\begin{equation}
\rho _{T}({\bf q})=\Omega ^{-1/2} \rho ({\bf q})+Zq\left(\frac{n\hbar}{2M
\omega_{TO}}\right)^{1/2} A({\bf q})\,,
\label{rhototal}
\end{equation}
where $A({\bf q})=a_{{\bf q}}+a_{-{\bf q}}^{\dag}$,
$M$ is the reduced mass of
the  basis ions, $Ze$ is the effective ionic charge,
$n$ is the electron density,
$\Omega$ is the system volume, and
the Fourier components of the electron density operator are
\begin{equation}
\rho ({\bf q}) = \sum_{i} e^{-i
{\bf q \cdot r}_{i}}=\sum _{{\bf k},\sigma} c^{\dag}_{{\bf k+q},\sigma}
 c_{{\bf k},\sigma}\,.
\end{equation}
The electron self-interaction
has been removed from $\tilde{H}_{g}$ by subtracting the term
with $N$, the number of electrons in the system.
The influence of interband electronic transitions is taken into account
by screening the Coulomb interaction $v_{q}$ with the high-frequency dielectric
constant $\epsilon _{\infty}$.

The effects of screening by conduction electrons and LO phonons are contained
in the interaction term in $\tilde{H}_{g}$ and will appear explicitly in the
total
dielectric function below.
The RPA or mean-field approximation for the interaction term is made by
working with
eigenstates $|n \rangle$
 of the Hamiltonian $H_{g}$ which neglects the terms in $\tilde{H}_
{g}$ that are quadratic in density fluctuations $\rho_{T}-\langle \rho_{T}
\rangle$\cite{ALLEN}.
\begin{eqnarray}
H_{g}|n \rangle &=& E_{n}|n \rangle \\
H_{g}&=&\tilde{H}_{g}-\frac{1}{2}\sum_{\bf q}v_{q}^{\infty}[\rho_{T}^{\dag}
({\bf q})-\langle\rho_{T}^{\dag}({\bf q})\rangle]
[\rho_{T}({\bf q})-\langle\rho_{T}({\bf q})\rangle]\,.
\end{eqnarray}
$H_{g}$ includes Coulomb interactions at the Hartree level.
Excitations
of the electron-phonon gas described by $H_{g}$
include both electron-hole pair and collective excitations.

Define
 $P(n,{\bf k} \rightarrow m,{\bf k}- {\bf q})$
 as the probability per unit time
for the probe electron to make a transition from $|{\bf k} \rangle$
to $|{\bf k}-{\bf q}  \rangle$ while the
electron-phonon gas  makes a transition from $|n \rangle$ to $|m \rangle$,
assuming that initially $|{\bf k} \rangle$ is occupied and $|{\bf k-q}
\rangle$ is unoccupied.
The interactions  included in $H_{g}$
produce charge density correlations, resulting in
an interaction potential between gas
excitations and probe electron which is screened
and assumed to
be weak. Using
Fermi's Golden rule (or the Born approximation)
with the probe particle plane wave states $|{\bf k}\rangle$,
the transition probability is
\begin{eqnarray}
P(n,{\bf k} \rightarrow m,{\bf k-q})&=&\frac{2 \pi }{\hbar}
\left(\frac{v_{q}^{\infty}}{\Omega }\right)^{2}
\mid \langle m \mid \rho _{T}^{\dagger}({\bf q}) \mid n \rangle \mid^{2}\,
 \delta (E_{n}-E_{m}+\hbar\omega_{k,k-q})
\label{Golden} \\
&=&\left(\frac{v_{q}^{\infty}}{\Omega \hbar}\right)^{2}
\int_{-\infty}^{\infty} dt e^{i\omega_{k,k-q}t}
\langle n \mid \rho_{T}({\bf q},t)\mid m \rangle \langle m \mid
\rho^{\dag} _{T}({\bf q},0) \mid n \rangle \\
\rho _{T} ({\bf q},t)&=&e^{iH_{g}t/\hbar} \rho _{T} ({\bf q})
e^{-iH_{g}t/\hbar}\,,
\end{eqnarray}
where $\omega_{k,k-q}=\hbar[k^{2}-({\bf k-q})^{2}]/2m^{*}$.

The inelastic differential scattering rate
$W^{inel}({\bf k,k-q})$
is the probability per unit time for an electronic transition from the
occupied state
$|{\bf k}\rangle$ to the unoccupied state $|{\bf k-q}\rangle$
due to scattering from the electron-phonon gas.
 It is found by summing
$P(n,{\bf k} \rightarrow m,{\bf k-q})$
over all possible final states $\mid m \rangle$ of the gas and averaging over
its initial states $\mid n \rangle$.
For the mobility problem, the average must be taken
over the {\em nonequilibrium} gas ensemble.

\subsection{Nonequilibrium ensemble averages}
Consider that an electric potential $U({\bf r},t)$ is applied
to the system of probe electron plus electron-phonon gas.
This influence is included by adding
the term
\begin{equation}
H_{ext}(t)= e \int d^{3}r\,U({\bf r},t)\rho_{T} ({\bf r},t)
\end{equation}
to the Hamiltonian of the system.
$H_{ext}$ is independent of time in steady state, but it is useful to require
the potential to be switched on at a time in the distant past when the gas
was in thermal equilibrium.
This requirement makes it possible to use the Kadanoff and Baym
method\cite{KAD2} for evaluating a nonequilibrium expectation value.
This method uses the equilibrium density matrix
to evaluate an ensemble average at a time after $U$ has been
switched on by using the
interaction picture to include
the dependence on $H_{ext}$ {\em explicitly}.
The nonequilibrium expectation value of an operator $ {\cal O}({\bf r},t)$
is\cite{KAD2}
\begin{eqnarray}
\langle {\cal O} ({\bf r},t)  \rangle _{noneq}&=&\langle {\cal O} _{U}
({\bf r},t) \rangle = {\cal Z}^{-1} {\rm Tr} \left\{ e^{-\beta H_{g}} {\cal O}
_{U}({\bf r},t) \right \} \label{exp}\\
{\cal O}_{U}({\bf r},t) &=& {\cal V} ^{-1} (t)\, {\cal O} ({\bf r},t)
\, {\cal  V} (t), \\
{\cal V} (t)&=&{\cal T} \exp \left[ \frac{-i}{\hbar} \int _{- \infty}^{t} dt'
H_{ext}(t') \right] \label{tdev}\,,
\end{eqnarray}
where
$\langle \,\, \rangle$ denotes an average over
the {\em equilibrium} population of the electron-phonon gas,
$\beta =1/k_{B}T$,
${\cal Z}={\rm Tr} \{ e^{-\beta H_{g}}\}$
is the partition function for the equilibrium gas,
and ${\cal T}$ denotes the time ordering operator which orders operators
with earliest times to the right.

In the
presence of $U({\bf r},t)$,
the transition rate
$P(n,{\bf k} \rightarrow m,{\bf k-q})$
in the Born approximation is unchanged from its equilibrium form
(\ref{Golden}) except that the influence of $H_{ext}$ must be
included in the time dependence of the matrix element.
Using equations (\ref{exp}-\ref{tdev})
to determine
$W^{inel}({\bf k,k-q})$ from
$P(n,{\bf k} \rightarrow m,{\bf k-q})$
for the nonequilibrium case yields
\begin{eqnarray}
W^{inel}({\bf k,k-q})
&=&\left(\frac{v_{q}^{\infty}}{\Omega \hbar }\right)^{2}
N S_{TU}({\bf q},\omega _{k,k-q})\,, \label{WUS} \\
S_{TU}({\bf q},\omega)&=&
\frac{1}{N}\int _{-\infty}^{\infty} dt \, e^{i\omega t} \langle
\rho _{TU}({\bf q},t) \, \rho _{TU}({\bf -q},0)\rangle \label{RHORHO}\\
\rho_{TU}({\bf q},t)&=&{\cal V}^{-1}(t)\,e^{iH_{g}t/\hbar}\,\rho_{T}\,
e^{-iH_{g}t/\hbar}\,{\cal V} (t), \label{rhoTUt}
\end{eqnarray}
where $S_{TU}({\bf q},\omega)$ is the nonequilibrium
total dynamic structure factor or spectral function for $\rho_{TU}$,
describing the
density fluctuation excitations of the nonequilibrium electron-phonon gas.
Similarly, the time-reversed rate is
\begin{equation}
W^{inel}({\bf k-q,k})
= \left(\frac{v_{q}^{\infty}}{\Omega \hbar }\right)^{2}
N S_{TU}(-{\bf q},-\omega _{k,k-q})\,. \label{WSU}
\end{equation}

When the gas excitations are not too highly damped so that they may usefully
be regarded as elementary excitations, and
when the deviations from equilibrium are slowly varying in space and time
on a scale determined by the frequencies and decay rates of the
elementary excitations, it is possible to define nonequilibrium distribution
functions and response functions\cite{KAD,DUB}.
The imaginary (dissipative) part of
the response function
for the nonequilibrium coupled electron-phonon system is
\begin{eqnarray}
{\rm Im}[\chi_{TU} ({\bf q},\omega )]
&=&\frac{-1}{2\hbar\Omega}\int _{-\infty}^{\infty}
dt  \, e^{i
\omega t} \langle  [ \rho_{TU} ({\bf q},t), \rho_{TU}
(-{\bf q},0)]
\rangle \label{IMCHITU} \\
&=&\frac{-n}{2\hbar}\left[S_{TU}({\bf q},\omega)
-S_{TU}(-{\bf q},-\omega)\right]
\label{noneqresponse}
\end{eqnarray}
where $[\rho , \rho ']$ is the commutator $\rho \rho' - \rho'\rho$.

In general, there is no simple relation between a nonequilibrium spectral
function and corresponding response function like
the fluctuation-dissipation
theorem (FDT) for the equilibrium case:
\begin{equation}
S_{T}({\bf q},\omega)=\frac{-2 \hbar}{n}\left[ N^{0}(\omega)+1\right]
{\rm Im}[\chi _{T}
({\bf q},\omega)] \,,\label{FDT}
\end{equation}
where $N^{0}(\omega) =(e^{\beta \hbar \omega}-1)^{-1}$
 is the equilibrium Bose-Einstein distribution function, while
$S_{T}$ and
${\rm Im}\chi_{T}$ are the {\em equilibrium} forms of
(\ref{RHORHO}) and (\ref{IMCHITU}).
In particular, since the relation between the time-reversed
nonequilibrium spectral functions
$S_{TU}({\bf q},\omega)$  and
$S_{TU}(-{\bf q},-\omega)$ is not known,
${\rm Im}[\chi_{TU} ({\bf q},\omega )]$
is not determined by
$S_{TU}({\bf q},\omega)$ alone.
It is possible, of course, to define
unknown dimensionless functions $N^{>}({\bf q},\omega)
$ and $N^{<}({\bf q},\omega)$ such that
\begin{eqnarray}
S_{TU}({\bf q},\omega)=\frac{-2\hbar}{n}N^{>}({\bf q},\omega){\rm Im}[\chi_{TU}
({\bf q},
\omega)]  \label{STU>}\\
S_{TU}(-{\bf q},-\omega)=\frac{-2\hbar}{n}N^{<}({\bf q},\omega){\rm
Im}[\chi_{TU}
({\bf q},
\omega)] \,. \label{STU<}
\end{eqnarray}
Equations (\ref{noneqresponse})and (\ref{STU>})-(\ref{STU<}) then imply
\begin{equation}
N^{>}({\bf q},\omega)-N^{<}({\bf q},\omega)=1\,. \label{unity}
\end{equation}
Adopting the notation $N^{<}({\bf q},\omega)=N({\bf q},\omega)$,
we have from (\ref{WUS}),(\ref{STU>}), and (\ref{unity}),
\begin{equation}
W^{inel}({\bf k,k-q})= \frac{-2 (v_{q}^{\infty})^{2}}
{\Omega \hbar}  \left[ N({\bf q},\omega _{k,k-q}) + 1 \right] \,
{\rm Im}[\chi_{TU}  ({\bf q}, \omega _{k,k-q})] \,. \label{WCHI}
\end{equation}

Equivalently, the Born aproximation for
 $W^{inel}$ can be expressed in terms of the
imaginary part of the total screened
Coulomb interaction $v_{q}/\epsilon_{TU}$ between the probe particle and the
electron-phonon gas,
\begin{equation}
W^{inel}({\bf k,k-q})
= \frac{-2v_{q}}{\Omega \hbar}
\left[ N({\bf q},\omega _{k,k-q}) + 1 \right]\,{\rm Im} \left[\epsilon
_{TU}^{-1}
({\bf q},\omega _{k,k-q})\right].
\label{WUN}
\end{equation}
This follows from the relations between the total
susceptibility $\chi_{TU}({\bf q},\omega)$, polarization
$P_{TU}({\bf q},\omega )$,
and dielectric function $\epsilon_{TU}({\bf q},\omega)$
for the nonequilibrium coupled system,
\begin{eqnarray}
\chi _{TU}({\bf q},\omega)=\frac{P_{TU}({\bf q},\omega)}{1-v_{q}^{\infty}
P_{TU}({\bf q},\omega)} \label{chi} \\
\epsilon_{TU}({\bf q},\omega)=\epsilon_{\infty}\left[1-v_{q}^{\infty}P_{TU}
({\bf q},\omega)\right]\,. \label{epstot}
\end{eqnarray}
The relation
$(v_{q}^{\infty})^{2}{\rm Im}\chi _{TU}({\bf q},\omega)=
v_{q}{\rm Im} \epsilon _{TU}^{-1}$
follows from (\ref{chi}) and (\ref{epstot}), and yields (\ref{WUN}) from
(\ref{WCHI}).

The weight factor $N({\bf q},\omega)$
in (\ref{WUN}) plays the role of the distribution
function for the nonequilibrium excitations of the coupled
electron-phonon system.
Just as for the equilibrium case in equation (\ref{FDT}), the structure
factor has been written as a product of a function (Im[$\epsilon^{-1}_{T}$])
describing the strength of interactions with the electron phonon gas, and a
function giving the occupation probabilities for the gas excitations.
However, it should be emphasized that
the FDT has not been used to derive (\ref{WUN}) and
$N({\bf q},\omega)$ is an {\em unknown} function.

\subsection{Separation of the total interaction}
The total
nonequilibrium structure factor $S_{TU}$
defined by (\ref{RHORHO}) is exactly separable into
a purely electronic part $S^{e}_{U}$ plus the remainder
$S_{U}^{ph}$ which
includes electron-phonon and ion-ion Coulomb interactions. Using the
nonequilibrium version of the total density operator
(\ref{rhototal}),
\begin{eqnarray}
S_{TU}({\bf q},\omega)&=&S^{e}_{U}({\bf q},\omega)+S^{ph}_{U}({\bf
q},\omega)\,,
 \label{STU} \\
S^{e}_{U}({\bf q},\omega)&=&\frac{1}{\Omega N}\int_{-\infty}^{\infty}dt\,
e^{i\omega t}\langle \rho_{U}({\bf q},t)\rho _{U}(-{\bf q},0)\rangle \,,\\
S^{ph}_{U}({\bf q},\omega)&=&\frac{1}{N}\int_{-\infty}^{\infty}dt\,e^{i\omega
t}
\left\{Zq\left(\frac{n\hbar}{2M\omega _{TO}
\Omega}\right)^{1/2}\Bigl[\langle \rho _{U}
({\bf q},t)A_{U}(-{\bf q},0)\rangle\Bigr.\right.\\
&+&\Bigl.\left.\langle A_{U}({\bf q},t)\rho_{U}(-{\bf
q},0) \rangle \Bigr] +(Zq)^{2}\frac{n\hbar}{2M\omega_{TO}}\langle A_{U}({\bf q}
,t)A_{U}(-{\bf q},0)\rangle \right\}\,.
\end{eqnarray}

This section relates $S^{e}$ and $S^{ph}$ to corresponding electron and
phonon parts of the response function, just as (\ref{STU>}) relates the total
structure factor to the total response function.
To do so, equation (\ref{WUN}) must be split into electron-electron and
electron-phonon parts.
Imposing the RPA for the dielectric function is a crucial step towards this
goal.
The response of the coupled system
in the RPA is the sum of the electron and ion response taken
separately\cite{ALLEN,VARGA}.
In this approximation, the total screened interaction $v_{q}/\epsilon_{TU}$
is the sum of a purely electron-electron part and an
electron-phonon interaction that includes a phonon self-energy
due to the polarization of the
conduction electrons\cite{DERIV}.
\begin{equation}
%% FOLLOWING LINE CANNOT BE BROKEN BEFORE 80 CHAR
%% FOLLOWING LINE CANNOT BE BROKEN BEFORE 80 CHAR
\frac{v_{q}}{\epsilon_{TU}}=\frac{v_{q}^{\infty}}{\epsilon_{U}}+\frac{M_{q}^{2}}
{\mid\epsilon_{U}\mid^{2}}D_{U}({\bf q},\omega)\,. \label{totint}
\end{equation}
The electron-electron part of the total interaction is $v_{q}^{\infty}/
\epsilon_{U}$, where $\epsilon_{U}$ is the nonequilibrium RPA dielectric
 function determined by electron-electron
interactions only. The polarization for the noninteracting electron gas
$P^{(1)}$
has the same
Lindhard form as the equilibrium case, except that the nonequilibrium electron
distribution function $f$ is used instead of the Fermi-Dirac
function\cite{DUB}.
\begin{eqnarray}
\epsilon_{U}({\bf q},\omega)
&=&1-v_{q}^{\infty}P^{(1)} ({\bf q},\omega)\label{RPA}\\
P^{(1)} ({\bf q},\omega)&=&\frac{1}{\Omega}\sum_{p,\sigma}\frac{f({\bf p})-
f({\bf p+q})}{\hbar\omega_{p,p+q}-\hbar\omega-i\delta}\,. \label{LINDHARD}
\end{eqnarray}
The remaining part  of the total interaction
is a product of the screened electron-phonon
matrix element $M_{q}^{2} |\epsilon_{U}|^{-2}$, and the
nonequilibrium phonon Green's function
$D_{U}({\bf q},\omega)$ containing the self-energy correction
$M_{q}^{2}D_{U}^{(0)}\chi_{U}$
to the noninteracting
Green's function $D_{U}^{(0)}$.
\begin{eqnarray}
M_{q}^{2}&=&v_{q}\frac{\hbar\omega_{LO}^{2}}{2\omega_{TO}}
\left(\frac{1}{\epsilon_{\infty}}-\frac{1}
{\epsilon_{0}}\right)\\
D_{U}({\bf q},\omega)&=&\frac{D_{U}^{(0)}}{1-M_{q}^{2}D_{U}^{(0)}\chi _{U}({\bf
q},\omega)}\,.
\end{eqnarray}

Now define unknown dimensionless functions
$N^{\stackrel{>}{<}}_{e}$ and $N^{\stackrel{>}{<}}_{ph}$ to relate the
electron and phonon components of $S_{TU}({\bf q},\omega)$ in (\ref{STU}) to
the
respective components of the imaginary part of $v_{q}\epsilon_{TU}^{-1}$ in
(\ref{totint}).
\begin{eqnarray}
(v_{q}^{\infty})^{2}S^{e}_{U}(\pm{\bf q},\pm \omega)= \frac{-2\hbar}{n}
v_{q}^{\infty}
N^{\stackrel
{>}{<}}_{e}({\bf q},\omega){\rm Im}[\epsilon^{-1}_{U}({\bf q},
\omega)] \label{Se}\\
(v_{q}^{\infty})^{2}S^{ph}_{U}(\pm{\bf q},\pm \omega)=\frac{-2\hbar}{n}
\frac{M_{q}^{2}}{\mid\epsilon _{U}\mid^{2}}
N^{\stackrel
{>}{<}}_{ph}({\bf q},\omega)
{\rm Im}[D_{U}({\bf q},\omega)]\,.\label{Sph}
\end{eqnarray}
Finally, using (\ref{WUS}),(\ref{STU}),(\ref{Se}), and (\ref{Sph}),
\begin{eqnarray}
W^{inel}({\bf k,k-q})=
\frac{-2}{\Omega \hbar}
\Biggl\{\Biggr. &v_{q}^{\infty}&
\left[ N_{e}({\bf q},\omega _{k,k-q}) + 1 \right]\,{\rm Im} \left[\epsilon
_{U}^{-1}
({\bf q},\omega _{k,k-q})\right] \nonumber \\
&+&\frac{M_{q}^{2}}{\mid\epsilon _{U}(q,\omega_{k,k-q})\mid^{2}}
\left[N_{ph}({\bf q},\omega _{k,k-q})+1\right]
{\rm Im} \left[ D_{U} ({\bf q},\omega_{k,k-q})\right]\Biggl.\Biggr\}\,.
\label{e+ph}
\end{eqnarray}
Within the RPA, the expressions
(\ref{WUN}) and (\ref{e+ph}) for $W^{inel}$ are equivalent. The last expression
contains {\em two} unknown functions, $N_{e}$ and $N_{ph}$, playing the roles
of the nonequilibrium occupation functions for the
excitations of the electronic and lattice
components of the coupled system.
Notice that $N_{e}$ must describe a distribution with boson properties.
This is because the density fluctuation $\rho_{q}^{\dag}$ produces excitations
of the electron gas which conserve the total number of electrons, such as
electron-hole pairs and plasmons, which have boson properties.

\section{Electron Boltzmann equation}
The time rate of change of the nonequilibrium electron distribution
 $f({\bf k})$ due to collisions
is determined
by the differential scattering rate
$W({\bf k,k-q})$; i.e., the probability
per unit time for the transition $|{\bf k}\rangle \rightarrow |{\bf k-q}
\rangle$. The depletion rate of $f({\bf k})$ is just $W({\bf k,k-q})$ weighted
by the probabilities that $|{\bf k}\rangle$ is occupied and $|{\bf k-q}\rangle$
is unoccupied, summed over all possible final states. The collision term,
$-\{\stackrel{.}{f}({\bf k})\}_{coll}$,
in the Boltzmann equation is this depletion rate minus the
replenishment rate determined similarly.
The electric field in the sample is ${\bf F({\bf r},t)}=-{\bf \nabla}
[U({\bf r},t)+\langle U_{ind}({\bf r},t)\rangle]$,
where $\langle U_{ind} \rangle$ is the average induced potential
resulting from the system's response to the applied potential $U$.
Thus, for a homogeneous system
under the influence of a static electric field ${\bf F}$,
\begin{eqnarray}
& &\frac{-e {\bf F}}{\hbar} \cdot \frac{\partial f({\bf k})}{\partial {\bf k}}
=-\left\{\stackrel{.}{f}({\bf k}) \right\} _{coll} \nonumber \\
&=&\frac{1}{\Omega}\sum_{{\bf q}}\Bigl\{ W({\bf k,k-q})f({\bf k})[1-f({\bf
k-q})]
-W({\bf k-q,k})f({\bf k-q})[1-f({\bf k})]\Bigr\}.
\label{fBOLT}
\end{eqnarray}
$W({\bf k,k-q})$
is a sum of rates for elastic and inelastic processes, $W=W^{el}+W^{inel}$.
Using the separation of
 the inelastic rate $W_{inel}$
in (\ref{e+ph}),
the inelastic contribution to the collision term separates into
electron-electron and electron-phonon collision terms,
\begin{equation}
\{\stackrel{.}{f}\}^{inel}_{coll}=\{\stackrel{.}{f}\}^{ee}_{coll}+
\{\stackrel{.}{f}\}
^{ep}_{coll}\,\,.
\end{equation}
\subsection{Electron-electron collision term}
The purely electronic component of $W^{inel}$ in (\ref{e+ph}) gives the
collision term
\begin{eqnarray}
-\{\stackrel{.}{f}({\bf k})\} ^{ee}_{coll} =& &
\nonumber \\
-\frac{2}{\hbar \Omega^{2}} \sum _{{\bf q}} v_{q}^{\infty}
\Biggl\{\Biggr.
[&N_{e}&({\bf q},\omega_{k,k-q})+1]
 {\rm Im}[\epsilon_{U}^{-1}({\bf q}, \omega_{k,k-q})]
 f({\bf k})
\left[1-f({\bf k-q})\right]
\nonumber \\   - &N_{e}&({\bf q},\omega_{k,k-q})
{\rm Im}[\epsilon_{U}^{-1}({\bf q},\omega _{k,k-q})]f({\bf k-q})
\left[ 1-f({\bf k})
\right] \Biggl.\biggr\}\,. \label{eecoll}
\end{eqnarray}
The next step towards putting the collision term into a useful form is to
eliminate the unknown function $N_{e}$, which gives the occupation
probabilities
of the density fluctuation excitations of the electron system.
This step is straightforward if an association is made between $N_{e}$ and
the nonequilibrum plasmon distribution, denoted here by $N_{p}$.
In steady state, there exists a
local detailed balance (LDB) between plasmons and electron-hole pairs which
can be used to eliminate $N_{e}$ in favor of $f$ in the
electron-electron collision
term\cite{DUB,WYLD}.
The LDB
condition is equivalent to the FDT for the nonequilibrium ensemble\cite{KLI}
which applies when $\chi$ determines the response not to external forces, but
to
internal fluctuation forces.
It holds when the Landau damping rate
$\gamma = 2{\rm Im} P^{(1)}$  is large compared to the time variation
of $f$, so that the plasmons
come into equilibrium with
the electrons {\em before} $f$ changes appreciably.
 For the mobility problem, $f$ is
stationary and the condition
always holds exactly, except at small $q$ values (where
$\gamma$ vanishes) which do not contribute to the electron collision integral
because of energy and momentum conservation restrictions.
If LDB is used in combination with the ``collisionless damping''
approximation\cite{DUB},
equation (\ref{eecoll}) is equivalent to the standard
single-particle electron-electron collision term, but screened with the
dynamic RPA dielectric function.
 This may be shown easily
for the case of weak damping, in which the plasmons are well-defined
excitations,
 by considering the
Boltzmann equation for the nonequilibrium
plasmon distribution $N_{p}(q)=N_{p}({\bf q},
\omega_{q})$
with $\omega_{q}=\omega_{p}(q)$\cite{WYLD,P&S}.
Plasmons are represented by the peaks in
the spectral function ${\rm Im} \epsilon^{-1}_{U}$. In the RPA, this function
contains  no corrections for collision broadening, so that the
relaxation of $N_{p}(q)$ is
completely described by coherent absorption and emission of
electron-hole pairs.
\begin{eqnarray}
\stackrel{.}{N}_{p}(q)=\frac{\pi v_{q}\omega_{q}}{\Omega^{2}}
\sum_{{\bf k}}\Biggl\{\Biggr.
[&N_{p}&(q)+1]
f({\bf k+q})\left[1-f({\bf k})\right]
\nonumber \\
-&N_{p}&(q)
f({\bf k})\left[1-f({\bf k+q})\right]\Biggl.\Biggr\}
\delta(\hbar \omega_{k,k+q}-\hbar \omega_{q})\,.\label{NBOLT}
\end{eqnarray}
In steady state,
$\stackrel{.}{N_{p}}=0$.
Therefore, the relation between the distributions $N_{p}$ and $f$ is
necessarily
\begin{eqnarray}
N_{p}(q){\rm Im}[P^{(1)}_{U}({\bf q},\omega_{q})]
=\frac{n}{2\hbar}S_{U}^{(1)}({\bf q},\omega_{q})
\label{LDB} \\
{\rm Im}[P^{(1)}_{U}({\bf q},\omega)]=\frac{2\pi}{\Omega}
\sum_{{\bf p}} \left[f({\bf p})-
f({\bf p+q})
\right]\, \delta (\hbar\omega_{p,p+q}-\hbar\omega) \label{IMCHI0} \\
S_{U}^{(1)}({\bf q},\omega)=\frac{4\pi\hbar}{N}\sum_{{\bf p}} f({\bf
p+q})\left[
1-f({\bf p})\right]\delta(\hbar \omega_{p,p+q}-\hbar\omega) \label{S0}\,,
\end{eqnarray}
relating the
 response
function ${\rm Im}[P^{(1)}_{U}]$ and structure factor $S^{(1)}_{U}$
for the noninteracting nonequilibrium electron gas.

The detailed balance relation between plasmons and electron-hole pairss
simplifies
the electron-electron collision term (\ref{eecoll}), if we assume
$N_{e}=N_{p}$.
Using the RPA interaction
${\rm Im}[\epsilon ^{-1}_{U}]
=v_{q}^{\infty}|\epsilon_{U}|^{-2}{\rm Im}[P^{(1)}_{U}]$
and (\ref{LDB}-\ref{S0})
gives
\begin{eqnarray}
-\{\stackrel{.}{f}({\bf k}_{1})\} ^{ee}_{coll}=
\frac{4\pi}{\hbar\Omega^{3}}
\sum _{2,3,4}
\frac{(v_{q}^{\infty})^{2}}
{|\epsilon_{U}(q,\frac{E_{1}-E_{3}}{\hbar})|^{2}}
 & \delta&(E_{1}+E_{2}-E_{3}-E_{4})
\delta ({\bf k}_{1}+{\bf k}_{2}-{\bf k}_{3}-{\bf k}_{4})
\nonumber \\
& \times & \Bigl\{ \Bigr. f({\bf k}_{1})f({\bf k}_{2})
[1-f({\bf k}_{3})][1-f({\bf k}_{4})]
 \nonumber \\
& & -
[1-f({\bf k}_{1})][1-f({\bf k}_{2})]f({\bf k}_{3})f({\bf k}_{4}) \Bigl.\Bigr\}
\,,
\label{eecoll2}
\end{eqnarray}
when the notational changes ${\bf k} \rightarrow {\bf k}_{1}, {\bf p}
\rightarrow {\bf k}_{2}$ are made, the summation variable ${\bf q}$ is
changed to ${\bf k}_{3}={\bf k}_{1}-{\bf q}$, and
a summation over a delta function for momentum conservation is included.
Notice that the nonequilibrium inverse dielectric function was needed to
obtain the collision term with nonequilibrium distributions for all four
(incoming and outgoing) electrons.
The remaining factor of $|\epsilon_{U}|^{2}$ in equation (\ref{eecoll2})
appears only in its equilibrium form $|\epsilon |^{2}$ if the Boltzmann
equation is linearized with respect to the field strength ${\bf F}$, since
the other factors vanish in equilibrium.
Also notice that $\epsilon_{\infty}$ rather than $\epsilon_{0}$ appears in
the electronic dielectric function $\epsilon_{U}(q,\omega)$ (see equation
(\ref{RPA})) which screens the interaction in the electron-electron
collision term (\ref{eecoll2}), unlike some previous
calculations using $\epsilon_{0}$ for
electron-electron scattering in GaAs\cite{LUG,CHATT}.

It should be emphasized that (\ref{eecoll2})
includes the effect of
electron-plasmon scattering through the resonance in $\epsilon (q,\omega)$
at $\omega=\omega_{q}$. For dc transport, the two-step process:
\begin{displaymath}
electron-hole\,\, pair \rightarrow plasmon \rightarrow electron-hole\,\, pair
\end{displaymath}
is equivalent to a dynamically screened electron-electron
scattering event because of the detailed balance between plasmons and
electrons. Since (\ref{eecoll2}) conserves total electron momentum
(in the absence of Umklapp processes),
its effect on the electrical current in materials with spherical energy
surfaces is nonvanishing only because it rearranges the momentum
distribution, which influences other scattering processes. The most important
difference between using (\ref{eecoll2}) and a previous
treatment\cite{FISC} of the effects of electron-plasmon scattering on mobility
is that the earlier work assumed an {\em equilibrium} (Bose-Einstein)
distribution function for the plasmons. In the latter case, there is a net
momentum loss from the nonequilibrium electron system.

\subsection{Electron-phonon collision term}
If the phonons
interacted only with electrons by non-Umklapp processes,
they would reach a local
detailed balance with the electrons just as the plasmons do.
But, in fact, the phonons interact with other phonons through anharmonic
lattice forces. The excess momentum given to the LO phonons by the
 nonequilibrium electron system is dissipated primarily through interactions
with acoustic phonons, which can lose momentum to the environment
through umklapp processes. Generally, mobility calculations have employed
the ``Bloch assumption'' that the phonon system may be treated as if it were
in thermal equilibrium\cite{ZIMAN}, valid when anharmonic processes are much
faster than electron-phonon scattering.

Adopting the model of equilibrium phonons,
the electron-phonon component of $W^{inel}$ in (\ref{e+ph})
gives the collision term,
\begin{eqnarray}
& & -\{\stackrel{.}{f}({\bf k})\} ^{ep}_{coll} =
\nonumber \\
& &-\frac{2}{\hbar \Omega^{2}} \sum _{{\bf q}}
\frac{M_{q}^{2}}{\mid\epsilon_{U}
(q,\omega_{k,k-q})\mid^{2}}
\Biggl\{ \Biggr.
\left[N^{0}(\omega_{k,k-q})+1\right]
 {\rm Im}D(q, \omega_{k,k-q})
 f({\bf k})
\left[1-f({\bf k-q})\right]
\nonumber \\ & &  - N^{0}(\omega_{k,k-q})
{\rm Im}D(q,\omega _{k,k-q})f({\bf k-q})
\left[ 1-f({\bf k})
\right] \Biggl.\Biggr\}\,. \label{epcoll}
\end{eqnarray}
The electron-phonon matrix element
is screened by only the electronic part of the total dynamic dielectric
function. The long-wavelength, static
limit of the RPA $\epsilon (q,\omega)$
is the temperature-dependent Thomas-Fermi dielectric
function with the {\em high-frequency} dielectric constant $\epsilon_{\infty}$.
Therefore, equation (\ref{epcoll}) in
this limit gives a somewhat
different result than treatments of electron-LO phonon scattering using
Thomas-Fermi screening with the
{\em static} dielectric
constant $\epsilon_{0}$\cite{LOW}.
For the nondegenerate case,
the long wavelength limit of (\ref{epcoll})
is in agreement with the original treatment of screening in
polar semiconductors by Ehrenreich\cite{EHREN}.

The ion-ion Coulomb interactions included in $\tilde{H}_{g}$ (equation
(\ref{Hg}))
 make a self-energy contribution
to the phonon Green's function\cite{MAHAN}. The real part of this contribution
shifts the LO phonon energy from $\hbar \omega_{TO}$ to
$\hbar \omega_{LO}$ in the denominator of
the noninteracting Green's function $D^{(0)}=-2\omega_{TO}/\hbar(\omega
_{LO}^{2}-\omega^{2})$. Damping due to anharmonic interactions and disorder
is neglected here.

The interacting phonon Green's function
\begin{equation}
D(q,\omega)=\frac{D^{(0)}}{1-M_{q}^{2}D^{(0)}\chi
(q,\omega)}=\frac{2\omega_{TO}/\hbar}
{\omega^{2}-\omega_{LO}^{2}-2\omega_{TO}M_{q}^{2}\chi(q,\omega)/\hbar}
\end{equation}
contains the self-energy term
$\Pi ^{ep} = M_{q}^{2}D^{(0)}\chi$ due to the polarization of the electron gas.
The real part of $\Pi^{ep}$ renormalizes the LO phonon frequency, giving rise
to two hybrid modes of mixed electron-phonon character.
The coupled modes have been treated previously\cite{PPSE}
with the
finite-temperature plasmon-pole approximation for $\chi$\cite{PP}.
In this approximation,
the electronic excitation spectrum which couples to the phonons is represented
by a single mode with energy varying between
$\hbar \omega _{p}=\sqrt{4\pi n e^{2}/m^{*}\epsilon_{\infty}}$
at small $q$ and $h^{2}q^{2}/2m^{*}$ at large $q$.
The plasmon coupling strength and energy are determined by requiring
fulfillment of the f-sum rule and zero-frequency Kramers-Kronig relation,
so that the plasmon-pole approximation for $\chi$ is\cite{PP,PPSE}
\begin{eqnarray}
\chi^{pp}(q,\omega)&=&\frac{1}{v_{q}^{\infty}}\frac{\omega_{p}^{2}}
{\omega^{2}-\tilde{\omega}_{p}^{2}}\\
\tilde{\omega}^{2}_{p}&=& \omega_{p}^{2}[1-\epsilon^{-1}(q,0)]^{-1}\,.
\end{eqnarray}
Using the plasmon-pole susceptibility $\chi^{pp}$ in the phonon self-energy
$\Pi^{ep}$ gives a
phonon Green's function with poles at the two frequencies $\omega_{+}$ and
$\omega_{-}$,
\begin{eqnarray}
D(q,\omega)&=&\frac{2\omega_{TO}(\omega^{2}-\tilde{\omega}^{2})}
{\hbar(\omega^{2}-\omega_{+}^{2})(\omega^{2}-\omega_{-}^{2})}\\
\omega_{\pm}^{2}&=&\frac{1}{2}\left\{\omega_{LO}^{2}+\tilde{\omega}^{2}_{p}\pm
\left[(\omega_{LO}^{2}-\tilde{\omega}^{2}_{p})^{2}+4\omega_{p}^{2}
(\omega_{LO}^{2}
-\omega_{TO}^{2})\right]^{1/2}\right\}\,. \label{OMEGAPM}
\end{eqnarray}

In the plasmon-pole model for the coupled modes, the phonon spectral
function $-\pi{\rm Im}[D(q,\omega)]$ appearing in the electron-phonon
collision term (\ref{epcoll})
has delta function peaks at $\omega_{+}$ and $\omega_{-}$,
\begin{eqnarray}
{\rm Im}[D(q,\omega)]={\rm Im}[D^{+}(q,\omega)+D^{-}(q,\omega)]\\
{\rm Im}[D^{\pm}(q,\omega)]=\mp\frac{\pi \omega_{TO}(\omega^{2}-\tilde{\omega}
_{p}^{2})}{\hbar\omega_{\pm}(\omega^{2}_{+}-\omega_{-}^{2})}[\delta(\omega+
\omega_{\pm})-\delta(\omega-\omega_{\pm})] \,.
\end{eqnarray}
At low densities, the low-energy hybrid mode has frequency $\omega{-}$ close
to the uncoupled plasmon frequency $\tilde{\omega}_{p}$, but has nonzero weight
in the phonon spectral function determined by the factor $(\omega_{-}^{2}-
\tilde{\omega}_{p}^{2})/(\omega_{+}^{2}-\omega_{-}^{2})$. Even though this
factor is small at low densities, the low-energy mode has an exponentially
larger thermal occupation $N^{0}(\omega_{-})$  at low temperatures than the
high-energy hybrid mode or uncoupled phonon mode do. Therefore, for low
densities and temperatures, the presence of $N^{0}$
in (\ref{epcoll}) suggests that the mobility calculated by including
scattering from the coupled modes should be lower than the mobility
calculated by using the uncoupled mode approximation.
In addition,
``antiscreening'' effects due to the dynamic nature of the system response
 can generate screened potentials that
are actually greater than in the unscreened case\cite{EHREN}.
Reference \cite{GAAS} presents numerical results for n-type GaAs
which show that
including dynamic screening, mode coupling, and electron-electron scattering
does significantly lower calculated mobilities
in situations where ionized impurity
scattering does not dominate, as in modulation-doped structures.

\section{Conclusion}
The total dielectric function $\epsilon_{T}(q,\omega)$ lends itself to a
simple and general method for deriving the inelastic collision term in the
electron Boltzmann equation for scattering from a coupled electron-phonon
system.
When the Born approximation is valid, the inelastic differential
scattering rate
$W^{inel}$ can be expressed in terms of the nonequilibrium total dielectric
function $\epsilon_{T}$, which includes screening by both electrons and
phonons.
In the RPA, $W^{inel}$ separates into a purely electron-electron interaction
plus an electron-phonon term describing interactions with the hybrid modes.
These interactions are
dynamically screened by only the
electrons. Reference \cite{GAAS} investigates the effects of the
phonon self-energy,
dynamical screening, and electron-electron scattering
on the mobility in n-type GaAs by means of a numerical
solution of the linearized Boltzmann equation.

Because the total dielectric function approach to the Boltzmann equation
treats the electrons
and ions in a unified manner, it can be applied quite generally to a
wide variety of coupled systems.
More accurate and complete
treatments\cite{PENN,ALLEN}
 of the lattice part of $\epsilon^{-1}_{T}$
could be useful
for more complex electron-phonon systems.
Also, it would be very interesting to try to include electron-impurity
scattering
through the $\epsilon_{T}(q,\omega)$ for an imperfect crystal with local
phonon modes.
Another logical
direction to pursue would start from a dielectric function which goes
beyond the RPA to include electron exchange effects in order to derive
the exchange scattering term in the Boltzmann equation. However, a systematic
approach to improving on the RPA is required, since exchange corrections can
be largely canceled by correlation terms.

\acknowledgements
I thank P.B. Allen, D.R. Penn, G.W. Bryant, G.D. Mahan, S. Das Sarma, and
J.R. Lowney for helpful conversations.


\begin{references}

\bibitem{KAD} L.P. Kadanoff and G. Baym, {\em Quantum Statistical Mechanics}
(Benjamin, Menlo Park, 1962).

\bibitem{HOL} T. Holstein, Ann. Phys. {\bf 29}, 410 (1964).

\bibitem{DUB} D.F. DuBois, in {\em Lectures in Theor. Physics}, Vol. IX C,
ed. W.E. Brittin (Gordon and Breach, New York, 1967) p.469.

\bibitem{SPI} V. \u{S}pi\u{c}ka and P. Lipavsk\'{y}, Phys. Rev. Lett. {\bf 73},
3439 (1994).

\bibitem{VAN} L. Van Hove, Phys. Rev. {\bf 95}, 249 (1954).

\bibitem{KIM} M.E. Kim, A. Das, and S.D. Senturia, Phys. Rev. B {\bf 18},
6890 (1978).

\bibitem{GAAS} B.A. Sanborn, ``Electron-electron interactions, coupled
plasmon-phonon modes, and mobility in n-type GaAs,'' submitted to Phys.
Rev. B.

\bibitem{PENN} D.R. Penn, S.P. Lewis, and M.L. Cohen, submitted to Phys. Rev.
B;
E.G. Maksimov, Zh. Eksp. Teor. Fiz. {\bf 69}, 2236 (1975); [Sov. Phys.-J.E.T.P.
{\bf 42}, 1138 (1976)].

\bibitem{ALLEN}
P.B. Allen, M.L. Cohen, and D.R. Penn, Phys. Rev. B. {\bf 38}, 2513 (1988).

\bibitem{MAHAN} G.D. Mahan {\em Many Particle Physics}, (Plenum, New York,
 1990), Section 6.3.

\bibitem{LUG} P.Lugli and D.K. Ferry, Physica {\bf 129B}, 532 (1985).

\bibitem{LOW} J.R. Lowney and H.S. Bennett, J. Appl. Phys. {\bf 69}, 7102
(1991);
W. Walukiewicz, L. Lagowski, L. Jastrzebski, M. Lichtensteiger,
and H.C. Gatos, J. Appl. Phys. {\bf 50}, 899 (1979); W. Walukiewicz,
J. Lagowski, and H.C. Gatos, J. Appl. Phys. {\bf 53}, 769 (1982).

\bibitem{CHATT} D. Chattopadhyay, J. Appl. Phys. {\bf 53}, 3330 (1982).

\bibitem{JAL} R. Jalabert and S. Das Sarma, Phys. Rev. B {\bf 41}, 3651 (1990).

\bibitem{DASSAR} S. Das Sarma, J.K. Jain, and R. Jalabert, Phys. Rev. B
{\bf 41}, 3561 (1990).

\bibitem{LEI} X.L. Lei and M.W. Wu, Phys. Rev. B {\bf 47}, 13338 (1993).

\bibitem{PP}
A.W. Overhauser, Phys. Rev. B {\bf 3}, 1888 (1971);
B.I. Lundqvist, Phys. Kondens. Mater. {\bf 6}, 193 (1967); {\bf 6}, 206 (1967).

\bibitem{PPSE} S. Das Sarma, J.K. Jain, and R. Jalabert, Phys. Rev. B
{\bf 37}, 4560 (1988); {\bf 37}, 6290 (1988).

\bibitem{KAD2} This method is explained in section 6-2 of reference \cite{KAD}.

\bibitem{VARGA} B.B. Varga, Phys. Rev. {\bf 137}, A1896 (1965).

\bibitem{DERIV}
The derivation given in reference \cite{MAHAN} for the equilibrium system
also holds for the nonequilibrium case.

\bibitem{WYLD} H.W. Wyld and D. Pines, Phys. Rev. {\bf 127}, 1851 (1962).

\bibitem{KLI} Y.L. Klimontovich, {\em The Kinetic Theory of Electromagnetic
Processes} (Springer-Verlag, Berlin, 1983);
Sov. Phys. JETP {\bf 48}, 852 (1979).

\bibitem{P&S}D. Pines and J.R. Schrieffer, Phys. Rev. {\bf 125}, 804 (1962).

\bibitem{FISC} M.V. Fischetti, Phys. Rev. B {\bf 44}, 5527 (1991).

\bibitem{ZIMAN} J.M. Ziman, {\em Electrons and Phonons} (Clarendon Press,
Oxford, 1960).

\bibitem{EHREN} H. Ehrenreich, J. Phys. Chem. Solids {\bf 8}, 130 (1959).

\end{references}
\end{document}